\documentstyle[12pt,preprint,prd,aps,floats,epsf]{revtex}

\def\calo{{\cal O}}
\def\call{{\cal L}}

\def\ie{{\it i.e.}}
\def\eg{{\it e.g.}}
\def\msusy{m_{\rm S}}
\def\be{\begin{equation}}
\def\ee{\end{equation}}
\def\bea{\begin{eqnarray}}
\def\eea{\end{eqnarray}}

\def\light{\phi}
\def\heavy{\Phi}

\begin{document}

\begin{titlepage}
\begin{center}
\hfill CERN--TH/95--114
\end{center}
\vskip .03in
\center{\large\bf Superfield Derivation
  of the Low-Energy Effective Theory of}
\center{\large\bf Softly Broken Supersymmetry}
 \vskip .6in
\center{{\sc Alex Pomarol and Savas Dimopoulos~\footnote{On leave of absence
from the Physics Department, Stanford University, Stanford CA 94305,
USA.}}}
\center{{\it Theory Division, CERN}\\{\it CH-1211 Geneva 23,
Switzerland}}
\vskip .3in
\begin{abstract}
We analyze the soft supersymmetry breaking parameters obtained in grand
unified theories after integrating out the heavy GUT-states. The superfield
formalism greatly simplifies the calculations and allows us to derive the
low-energy effective theory in the general case of non-universal and
non-proportional soft terms by means of a few Feynman diagrams.
We find new contributions not considered before. We discuss the
implications for the destabilization of the gauge hierarchy, flavor
violating processes and the non-unification of sparticle masses in
unified theories.
\vskip .3in
\noindent CERN--TH/95--114\hfill\\
\noindent May,$\,\,$1995\hfill\
\end{abstract}
\end{titlepage}

\section{Introduction}
\label{sec:s1}

The idea of gauge unification in the minimal supersymmetric model
 (MSSM)\cite{su5a,su5b}
seems to be supported by the recent experimental data \cite{langacker}.
The  unification scale
 $M_G\approx 2\times 10^{16}$ GeV is just below the Planck scale
 $M_P\approx 2\times 10^{18}$ GeV
but much  higher than the weak scale $M_Z$.
Supersymmetry plays the role of
stabilizing  this gauge  hierarchy against radiative corrections. Even
if supersymmetry is  broken, the hierarchy can be kept  stable
if the breaking  arises  only from soft terms
 (these are terms that
do not reintroduce quadratic divergencies) \cite{su5a,girardello}:
\begin{equation}
-{\cal L}_{soft}=
m^2_{ij}\phi^*_i\phi_j+\left(\frac{1}{6}A_{ijk}\phi_i\phi_j
\phi_k+\frac{1}{2}B_{ij}\phi_i\phi_j+\frac{1}{2}
\widetilde{M}_{i}\tilde\lambda_i^2+h.c.\right)\, ,
\label{softterms}
\end{equation}
where $\phi_i(\tilde\lambda_i)$ are the scalar (gaugino)
fields  of the
theory.
Eq.~(\ref{softterms}) parametrizes the most general
soft supersymmetry breaking (SSB) terms.
In order to maintain the gauge hierarchy,  the
scale of   supersymmetry breaking, $\msusy$, must be close to
$M_Z$.

In  supergravity theories, where
supersymmetry is assumed to be broken in a hidden sector
which couples only gravitationally to the observable sector,
the SSB terms  of eq.~(\ref{softterms})
are generated at the Planck scale \cite{sugra,hall}.
In some  supergravity models,  one can obtain relations between
 the SSB parameters and
then reduce the number of independent parameters. For example,
in minimal supergravity theories where the K\"ahler potential is flat,
 one finds that   the SSB parameters have universal values at
$M_P$ \cite{sugra,hall}, {\it i.e.},
\be
m^2_{ij}\equiv m^2_0,\ \ B_{ij}\equiv B_0M_{ij},\ \
A_{ijk}\equiv A_0Y_{ijk},\ \  \widetilde{M}_{i}\equiv \widetilde{M}_{1/2}\, ,
\label{softuni}
\ee
where $Y_{ijk}(M_{ij})$ are the trilinear (bilinear)
couplings in the superpotential.
Other examples in which  relations between the SSB parameters can be derived,
can be found in
 superstring theories \cite{string}.

At lower
energy scales, however,
the SSB parameters deviate from their initial values
at $M_P$
 according to
 the renormalization
group equations (RGEs) of the corresponding effective theory.
In  grand unified theories (GUTs),
 the SSB parameters will evolve from $M_P$ to $M_G$
according to the RGEs of the GUT. At $M_G$, one has to integrate out the
heavy particles [of masses of $\calo(M_G)$] and
evolve again the SSB parameters
 from $M_G$ to $M_Z\sim\msusy$ with the RGEs of the MSSM.
In these two  processes,  running and integrating out,
the SSB parameters  can be shifted from their initial  values
at $M_P$.
Since the low-energy sparticle spectrum depend on the SSB parameters,
the study of these effects is crucial
for the  phenomenology of the MSSM.
Nevertheless, to compute the effects one has to specify
the GUT, rendering the studies  very model-dependent.
Partial analysis
can be found in
 refs.~\cite{hall},\cite{hkr}-\cite{dterms}.

The purpose of this  paper is to carry out a general study of
how the SSB parameters are modified when the heavy particles
 of a GUT or a flavor theory at a high scale $M_G$,
 are integrated out at tree-level.
We will consider the most general softly
broken  supersymmetric theory and  will use superfield techniques.
 Previous  analysis \cite{hall,sen,giudice,murayama}
have been carried out in  component fields instead of superfields,
 requiring lengthy calculations. Moreover, conditions
such as
universality \cite{hall,giudice}  or proportionality \cite{sen,murayama}
have been  assumed in order to simplify the calculations.
Here we will reproduce these previous results
in an easier way using superfield techniques,
generalize them  and further
pursue their phenomenological implications.
As we will see,  integrating out the heavy modes can
be easily accomplished by Feynman diagrams.

Softly broken supersymmetric theories can be formulated in the superfield
formalism by using a spurion external field, $\eta$\cite{girardello}.
 Supersymmetry
is broken by giving to this superfield a $\theta$-dependent value,
$\eta\equiv\msusy\theta^2$. Then,
the most general
Lagrangian describing a softly broken supersymmetric theory
can be written as\footnote{
We do not  include terms with only
gauge vector-superfields since they are not relevant for our analysis.
We  assume  that
the gauge group of the GUT is simple and that
the chiral superfields
are non-singlets under the GUT-group \cite{bagger}.
 Our notation and  conventions
 follow ref.~\cite{wess}.}
\begin{equation}
{\cal L}={\cal L}_{SUSY}+{\cal L}_{soft}
\end{equation}
where
\begin{eqnarray}
{\cal L}_{SUSY}&=&\int d^4\theta{\bf \Phi}^{\dagger} e^{2gV}{\bf \Phi}
+\left(\int d^2\theta W({\bf \Phi})+h.c.\right)\, ,\label{susypart}\\
{\cal L}_{soft}&=&\int d^4\theta{\bf
 \Phi}^{\dagger}[\bar\eta\Gamma^*+\eta\Gamma-
\bar\eta\eta Z]e^{2gV}{\bf \Phi}-\left(\int d^4\theta\bar\eta\eta{\bf \Phi}^T
\frac{\Lambda}{2}{\bf \Phi}
+\int d^2\theta \eta W'({\bf \Phi})+h.c.\right)\, ,\label{softsf}
\end{eqnarray}
where the column vector ${\bf \Phi}=({\bf \Phi}_1,{\bf \Phi}_2,...)^T$ denotes
 the
 chiral superfields
of the theory, $V\equiv T_AV_A$
are  the vector superfields
and the matrices  $\Gamma$, $Z$  and $\Lambda$ denote
dimensionless SSB parameters.
Eq.~(\ref{softsf}) leads, in  component fields [and after
replacing $\eta(\bar\eta$) by $\msusy\theta^2(\bar\theta^2$)], to
 eq.~(\ref{softterms}) with
\begin{eqnarray}
A_{ijk}&=&\msusy\left[Y_{ijk}'+Y_{ljk}\Gamma_{li}+Y_{ilk}\Gamma_{lj}
+Y_{ijl}\Gamma_{lk}\right]\, ,\nonumber\\
B_{ij}&=&\msusy\left[M_{ij}'+M_{lj}\Gamma_{li}+M_{il}\Gamma_{lj}
+\msusy\Lambda_{ij}\right]\, ,\nonumber\\
m^2_{ij}&=&\msusy^2\left[Z_{ij}+\Gamma^{*}_{il}\Gamma_{lj}\right]\, ,
\label{SSBcomp}
\end{eqnarray}
where $Y_{ijk}'(M_{ij}')$ are the trilinear (bilinear)
couplings in $W'({\bf \Phi})$.
This function $W'({\bf \Phi})$
is in principle a general holomorphic function of the
superfields different from the superpotential $W({\bf \Phi})$.
In supergravity theories, however,
where supersymmetry is broken by a hidden sector that does
not  couple to the observable sector in the superpotential,
one has that \cite{hidden}
\be
W'({\bf \Phi})=aW({\bf \Phi})\, ,
\label{proport}
\ee
where $a$ is a constant. Eq.~(\ref{proport}) will be
referred as
the  $W$-proportionality condition.
It is crucial to note that $\int d^2\theta\eta W'$
 does not renormalize. It is due to the
non-renormalization theorem which states that
terms under the integral $\int d^2\theta $ do not receive radiative
corrections \cite{grisaru}.
Then, if the  proportionality of $W'$ to $W$
 is satisfied at the scale
where supersymmetry is broken, $\it i.e.$, Planck scale in supergravity,
 it will
be satisfied at any lower scale.

{}From eq.~(\ref{SSBcomp}), we see that
 the  universal conditions  for $m^2_{ij}$ and $A_{ijk}$ [eq.~(\ref{softuni})]
are satisfied when  eq.~(\ref{proport}) holds and
\begin{equation}
\Gamma\ ,\ Z\propto {\bf 1}.
\label{universal}
\end{equation}
Nevertheless, even if eq.~(\ref{universal})  holds at the  scale
where the SSB terms are generated ($\sim M_P$),
 renormalization
effects  modify  $\Gamma$ and $Z$.
(The  renormalization
 of $\Gamma$ and $Z$ can be found in ref.~\cite{yamada}.)
 At $M_G$, deviations from universality
can be sizeable \cite{pre,nir,barbieri}.

One of the main motivations for  the analysis
in refs.~\cite{hall,sen,giudice,murayama}
was to see whether integrating out the heavy GUT-modes, generates
 SSB terms of $\calo(M_G\msusy)$ in the low-energy theory
that could  destabilize the gauge hierarchy.
It was shown in refs.~\cite{sen,murayama} that in fact such terms
can be present if the SSB parameters are non-universal.
Here we will show that  the superfield formalism
allows us to understand easily
the effects that destabilize the gauge hierarchy.

The effects of integrating out the heavy GUT-modes
are also very important
for phenomenological purposes.
They  modify the SSB parameters and consequently the
sparticle spectrum.
Two types of effects in the sparticle masses are of special interest.
Effects that lead to FCNC \cite{hkr,barbieri} and
 effects that modify mass GUT-relations \cite{murayama,us,murayamab,dterms}:

\noindent{\bf 1. FCNC (Horizontal effects):}
For arbitrary values  of $\sim\calo(1$ TeV) for $m_{ij}$ and $A_{ijk}$,
the squark and slepton contribution to FCNC processes typically exceed
the experimental bounds \cite{su5a}.
These processes put
severe constraints on the masses of
the first and second
family of squarks and sleptons  \cite{fcnc}.
These constraints can be satisfied if we demand
\begin{quote}
 ({\it a}) universal  soft masses for  the squarks and sleptons
 [condition
 (\ref{universal})]\cite{su5a},

 ({\it b}) proportionality
between  $A$-terms
and the corresponding Yukawa-terms, \ie, $A_{ijk}\propto Y_{ijk}$
 [$W$-proportionality and condition (\ref{universal})].
\end{quote}
Grand unified  models which as a result of
 integrating out the heavy GUT-modes, generate  large deviations from
 (\ref{proport}) and (\ref{universal}) for the squarks and sleptons
will not lead to  viable low-energy theories.
A priori, it seems  that this is the case
 in GUTs  or flavor models  that have
the  different families of quarks and leptons couple
to different Higgs representation above $M_G$.
Since the {\it gauge} renormalization of the
the trilinear $A$-terms depend on the GUT-representation of the
fields, these corrections
will not be universal in flavor space.
We will show, however, that once the heavy states are integrated out,
such effects cancel out from the
low-energy effective theory.

\noindent{\bf 2. Unification of soft masses (Vertical effects):}
In GUTs,
where
  quarks and leptons are embedded in fewer
 multiplets, one expects that,   due to the GUT-symmetry,
 the number of independent SSB parameters is reduced.
 For example, in the minimal SU(5)
one has
\begin{equation}
\{Q,U,E\}\, \in\ {\bf 10}\ ,\
\{L,D\}\, \in\ \bar{\bf 5}\, ,
\end{equation}
where $Q$  ($L$)
and $U$, $D$ ($E$) are respectively
  the quark  (lepton) SU(2)$_L$-doublet and singlets.
Thus, one expects at $M_G$
\be
m^2_{Q}=m^2_{U}=m^2_{E}\equiv m^2_{10}\ ,
\ m^2_{L}=m^2_{D}\equiv m^2_{5}\, .
\label{GUTrelation}
\ee
Nevertheless, such GUT-relations
can be modified by GUT-effects even at tree level.
When  the heavy GUT-modes are integrated out,
the  soft masses of the sparticles that belong
to the same GUT-multiplet
can be  split. This is known to happen, for example,
 in SU(5) theories \cite{us}
if  more than three families are present at $M_G$ or in
SO(10) theories where $D$-term contributions split
the soft masses of the matter fields in the {\bf 16}
 representation\cite{dterms}.

In section II  we will calculate  in the superfield formalism
 the  shifts in the SSB parameters of the low-energy theory
induced  when  heavy modes
 are integrated out.
We will consider non-universal SSB parameters
($\Gamma$ and $Z$ will be arbitrary matrices).
We will first
 assume  $W$-proportionality  (sections IIA and IIB),
and subsequently
will analyze the effects of relaxing this assumption  in section IIC.
In section III we will  study the phenomenological implications focusing
on the stability of the gauge hierarchy,
 FCNC and  the unification of the soft masses at $M_G$.
In section IV we will present our conclusions.

\section{Integrating out the heavy superfields}

Eqs.~(\ref{susypart}) and (\ref{softsf}) parametrize the most general
softly broken supersymmetric theory above $M_G$.
 At $M_G$, some of
the chiral superfields get  vacuum expectation values (VEVs)
 breaking the gauge symmetry of the GUT to
 SU(3)$\times$ SU(2)$_L\times$U(1)$_Y$,  the MSSM group.
We can rotate the superfield ${\bf \Phi}-\langle{\bf \Phi}\rangle$ to
a basis where the supersymmetric mass matrix is diagonal (SUSY-physical
basis) and
separate the superfields in lights and heavies:
\be
{\bf \Phi}\rightarrow{\bf \Phi}'=\langle\heavy \rangle+\heavy
+\light\equiv\left(
\matrix{
\langle\heavy_i\rangle\cr 0}\right)+
\left(\matrix{\heavy_i\cr 0}\right)+
\left(\matrix{0\cr \light_\alpha}\right)\, ,
\label{separation}
\ee
where $\langle\heavy \rangle$ is of  ${\cal O}(M_G)$ and\footnote{
Greek (latin) letters denote light (heavy) superfields.}

\bea
\heavy_i&:&\ {\rm Heavy\ superfields\ of\
mass\ (but\ no\ VEV)\ of}\ {\cal O}(M_G)\, ,\nonumber\\
\light_\alpha&:&\ {\rm Light\ superfields\ of\ mass\ of}\
 {\cal O}(\lesssim\msusy)\, .
\eea
Assuming $W$-proportionality
[eq.~(\ref{proport})], the SSB terms (\ref{softsf})
 are  given by
\begin{eqnarray}
{\cal L}_{soft}=\int d^4\theta&\Bigg\{&\langle\heavy^{\dagger} \rangle
[\bar\eta\Gamma^*+\eta\Gamma-
\bar\eta\eta Z]e^{2gV}\langle\heavy\rangle\nonumber\\
&+&\left\{\langle\heavy^{\dagger} \rangle
[\bar\eta\Gamma^*+\eta\Gamma-
\bar\eta\eta Z]e^{2gV}\heavy
-\langle\heavy^T \rangle\bar\eta\eta\Lambda\heavy
 +h.c.\right\}\nonumber\\
&+&(\heavy^{\dagger}+\light^{\dagger})
[\bar\eta\Gamma^*+\eta\Gamma-
\bar\eta\eta Z]e^{2gV}(\heavy +\light)\nonumber\\
&-&\left\{(\heavy^T +\light^T)
\bar\eta\eta\frac{\Lambda}{2}(\heavy +\light)
+h.c.\right\}\Bigg\}\label{extraterm}\\
-\Bigg(\int d^2\theta&\eta a&W+h.c.\Bigg)\, ,\label{propor}
\end{eqnarray}
where now
 $\Gamma$, $Z$ and $\Lambda$
define the SSB parameters in the new basis (\ref{separation}). In this basis,
it is obvious that the superpotential does not contain either
 mass terms mixing
heavy with light superfields
or
 linear terms
 with the heavy superfields.
Since  we assumed $W$-proportionality
  at $M_P$,
 the non-renormalization theorem guarantees that such terms are also absent
 in eq.~(\ref{propor}) at $M_G$.
We have also assumed that there are not light MSSM-singlets
in the model. We can easily see
that if light singlets are present, a term
\begin{equation}
\int d^4\theta\langle\heavy^{\dagger} \rangle
\bar\eta\Gamma^*\light\ \ \sim\ \ M_G\msusy\left(\frac{\partial W}
{\partial \light}\right)^*\, ,
\label{lights}
\end{equation}
can be induced and spoil the gauge hierarchy\footnote{
There are different possibilities to suppress eq.~(\ref{lights})
and allow light singlets\cite{nilles}.
We will not consider such alternatives here.}\cite{nilles}.

We are now ready to integrate out the heavy superfields.
We will first consider the heavy chiral-superfields
($g=0$), leaving the gauge sector
for later.

\subsection{Integrating out the heavy chiral-sector}

To integrate out the heavy chiral-superfields at tree-level,
 we compute their
 equations of motion
and use them
to write the heavy superfields
as a function of the light and spurion superfields, $\ie$,
$\heavy_i =f(\light_\alpha,\eta)$.
When one derives the equations of motion by the variational principle,
one must take into account the fact that the chiral
superfields are subject to the constraints $\bar D\Phi=0$ where
 $D=\partial_{\theta}+i(\sigma^\mu\bar\theta)\partial_\mu$
is
the supersymmetric covariant derivative.  However,
using  the relation
\be
 \int d^4\theta F(\Phi,\Phi^{\dagger})
=-\int d^2\theta\frac{\bar D^2}{4}F(\Phi,\Phi^{\dagger})\, ,
\label{reverse}
\ee
one  can write the action as an integral over the chiral
superspace ($\int d^2\theta$) where the chiral superfields
 are unconstrained and calculate the equations of motion
by the variational principle\cite{sohnius}.
Hence
\be
\frac{\delta}{\delta\heavy_j}\Bigg\{\int d^2\theta
W(\heavy,\light) (1-a\eta)
-\int d^2\theta\frac{\bar D^2\bar\eta}{4}
\bigg(\langle\heavy^{\dagger} \rangle
\left[\Gamma^*
-\eta Z\right]\heavy
-\langle\heavy^T \rangle\eta\Lambda\heavy\bigg)+\dots\Bigg\}=0\, ,
\label{equationm}
\ee
where
\be
W(\heavy,\light)=\frac{1}{2}
M_{ij}\heavy_i\heavy_j
+\frac{1}{2}Y_{i\alpha\beta}\heavy_i\light_\alpha\light_\beta+
\dots
\ee
is the superpotential in the SUSY-physical basis (\ref{separation}).
We have  kept not only the dominant terms of
 $\calo(M_G)$  but also  terms of $\calo(1)$ that
involve  a heavy superfield and two light superfields.
It will become clear later, that the terms neglected do not lead to
any contribution in the limit $M_G\gg\msusy$.
 Eq.~(\ref{equationm}) leads to
\begin{equation}
\heavy_i\simeq\frac{-1}{M_{ij}}\left[
\frac{1}{2}Y_{j\alpha\beta}\light_\alpha\light_\beta
-\frac{\bar D^2\bar\eta}{4(1-a\eta)}
\left\{\langle\heavy^{\dagger} \rangle\left[\Gamma^*-\eta Z\right]
-\langle\heavy^T \rangle\eta\Lambda\right\}_j\right]\, .
\label{eoms}
\end{equation}
Eliminating the heavy superfields from the effective Lagrangian
through the above equations of motion,
and using eq.~(\ref{reverse}), we get
\begin{equation}
{\cal L}_{eff}(\light,\eta)={\cal L}(\heavy =0)-
\int d^4\theta
\left\{\langle\heavy^{\dagger} \rangle\left[\bar\eta\Gamma^*-\bar\eta\eta
Z\right]-\langle\heavy^T \rangle\bar\eta\eta\Lambda\right\}_j
\frac{Y_{i\alpha\beta}}{2M_{ij}}
\light_\alpha\light_\beta+h.c.\, .
\label{muterm}
\end{equation}
Replacing $\bar\eta\rightarrow\msusy\bar\theta^2$, the
second term of the r.h.s. of
eq.~(\ref{muterm}) gives a mass term of $\calo(\msusy)$
to  the superpotential of the light superfields
 (a $\mu$-term  for
 the light Higgs doublets of the MSSM can be induced in this way
\cite{hall}).
Note, however, that  this term
 is only generated if there is  a trilinear
coupling
between a heavy MSSM-singlet and two light superfields.

Had we worked in component fields instead of superfields, the process
of integrating out the heavy modes would have been
 much more complicated since
we have to deal with the scalar fields and auxiliary fields independently
with complex equations of motion \cite{hall,sen,giudice,murayama}.
Eq.~(\ref{muterm})  gives the result of ref.~\cite{hall}
obtained here in a much simpler way using superfields techniques.
Furthermore, it is valid for a more general class of theories,
since universality is not assumed.

A set of simple rules to obtain
the terms (\ref{muterm})
  can  be easily derived.  These rules are:
\begin{quote}
({\it i})
Draw all possible Feynman diagrams with
  heavy superfields in the internal lines and
light and spurion superfields in the external lines.

({\it ii}) For each external line, write the corresponding
superfield $\light_\alpha$, $\eta$ (or   $\light^\dagger_\alpha$, $\bar\eta$).

({\it iii}) For each  $\langle\heavy_i \heavy_j \rangle$
 propagator,  write ${-1}/{M_{ij}}$.

({\it iv}) Vertices are read directly from the Lagrangian
eqs.~(\ref{susypart}), (\ref{extraterm}) and (\ref{propor}).

({\it v}) Integrate over $\int d^4\theta$ ($\int d^2\theta$)
if at least one vertex (none of the vertices) comes from a $D$-term.
\end{quote}
Following the rules above, we have that the only
diagrams that do not
go to zero in the heavy limit $M_{ij}\gg\msusy$ are those given in
 fig.~1  and they give
 the contribution  eq.~(\ref{muterm}).

\subsection{Integrating out the heavy gauge vector-sector}

If the gauge symmetry of the GUT is broken by the VEVs of the
 chiral superfields, the vector superfields
associated with the broken generators
get a mass term  of $\calo (M_G^2)$:
\begin{equation}
2g^2\int d^4\theta \langle\heavy^{\dagger} \rangle
T_AT_B\langle\heavy \rangle V_AV_B\, .
\label{massvector}
\end{equation}
One can perform a rotation in the $V_A$ and work in a basis where
the mass matrix of the heavy vector-superfield is diagonal.
In the supersymmetric limit, each broken generator has a chiral
superfield associated, $\langle\heavy^{\dagger} \rangle T_A\heavy$,
that contains the Goldstone boson and its
superpartners. We will work in the super-unitary gauge \cite{fayet}
where these chiral superfields have been gauged away:
\begin{equation}
\langle\heavy^{\dagger} \rangle T_A\heavy =0\, .
\label{unitaryg}
\end{equation}
Expanding the exponentials in eqs.~(\ref{susypart}) and
 (\ref{extraterm})
and using eqs.~(\ref{separation}), (\ref{unitaryg})
and the condition that supersymmetry is not broken
by  the observable sector (that the $D$-terms do not get VEVs),
\be
\langle\heavy^{\dagger} \rangle T_A\langle\heavy \rangle =0\, ,
\label{dterms}
\ee
we have
\bea
{\cal L}_{SUSY}=2\int d^4\theta&\Bigg\{&\light^{\dagger} gT_A\light V_A
+\left[\frac{1}{2}\light^{\dagger} g^2T_AT_B\light+
\langle\heavy^\dagger\rangle g^2T_AT_B(\heavy+\light)+h.c.\right]V_AV_B
\nonumber\\
&+&\frac{1}{2}M^2_AV_A^2
+\frac{2}{3}\langle\heavy^{\dagger}\rangle
 g^3T_AT_BT_C\langle\heavy\rangle
V_AV_BV_C\Bigg\}+\dots\, ,\nonumber\\
{\cal L}_{soft}=2\int d^4\theta&\Bigg\{&V_A\eta\bigg[
\langle\heavy^{\dagger} \rangle gT_A\Gamma\langle\heavy \rangle
+\langle\heavy^{\dagger}
 \rangle gT_A\Gamma(\heavy +\light)\nonumber\\
&+&(\heavy^{\dagger} +\light^{\dagger})
gT_A\Gamma\langle\heavy \rangle +
\light^{\dagger} gT_A\Gamma\light\bigg]+
\eta\langle\heavy^{\dagger} \rangle  g^2T_AT_B\Gamma
\langle\heavy \rangle V_AV_B+h.c.\Bigg\}\nonumber\\
-2\int d^4\theta &V_A&\eta\bar\eta\langle\heavy^{\dagger} \rangle
gT_AZ\langle\heavy \rangle +\dots\, ,
\label{vector}
\eea
where $M^2_A=2g^2\langle\heavy^{\dagger} \rangle
T^2_A\langle\heavy \rangle$. It will be justified a posteriori,
when we  use Feynman diagrams to
integrate out the
heavy vector superfields, why
only the  terms kept
 in eq.~(\ref{vector}) are relevant for the calculation.
Gauge invariance implies $[T_A,\Gamma]=[T_A,Z]=0$.
It is easy to check using eqs.~(\ref{unitaryg})  and
(\ref{dterms}) that
if  $\Gamma,Z\propto {\bf 1}$,
the equations of motion for the vector superfields are simple
$V_A=0+\calo(1/M^2_G)$ and they decouple from the effective theory
in agreement with ref.~\cite{hall}.  For general SSB parameters,
 this will not be the case.

Let us first consider the case where the $V_A$ are not singlets
under the MSSM gauge group. This is always the case if the rank of the
GUT-group is not larger than the rank of the MSSM group ({\it e.g.},
SU(5)).
One then has that the linear terms
 with  $V_A$  in (\ref{vector}) vanish. Thus,
the equation of motion is given by
\begin{equation}
V_A=\frac{-g}{M^2_A}\left(\eta
\langle\heavy^{\dagger} \rangle
T_A\Gamma\light+\eta\light^{\dagger} T_A\Gamma\langle\heavy \rangle
+h.c.\right)+\calo(\frac{1}{M_G^2})\, ,
\label{vectoreom}
\end{equation}
that inserting it back in  eq.~(\ref{vector}) gives  a new effective term
\begin{equation}
{\cal L}_{eff}'=
-\int d^4\theta\frac{g^2}{M^2_A}
\left(\eta\langle\heavy^{\dagger} \rangle T_A\Gamma\light+\eta
\light^{\dagger} T_A\Gamma\langle\heavy \rangle+h.c.\right)^2\, .
\label{induced}
\end{equation}
The above term can be  easily obtained by
applying the rules ({\it i})--({\it v}) with
\begin{quote}
\noindent ({\it vi})   For each  $\langle V_AV_A\rangle$
 propagator,  write $-1/{(2M_{A}^2)}$.
\end{quote}
The only possible Feynman diagram  is given in fig.~2 and gives the
contribution obtained in eq.~(\ref{induced}).

Finally, let us consider the  case where $V_A$ can also be a MSSM-singlet.
Applying the rules ({\it i})--({\it vi}), we have the Feynman
 diagrams of fig.~3 that  give the new contributions
\bea
\call_{eff}''&=&
-2\int d^4\theta\left(\eta\light^{\dagger} gT_A\light
 K_A+h.c.\right)\label{triterm}\\
&-&2\int d^4\theta\bar\eta\eta\Bigg\{-\{\langle\heavy^{\dagger}\rangle
[gT_A(\Gamma^*K_A+h.c.)-g^2\{T_A,T_B\}K_AK_B^*]\}_j
\frac{Y_{i\alpha\beta}}{2M_{ij}}
\light_\alpha\light_\beta\nonumber\\
&+&\light^{\dagger} gT_A\light
\left(\frac{\langle\heavy^{\dagger} \rangle
g^3(T_AT_BT_C)_{sym}\langle\heavy \rangle K_BK^*_C}{3M^2_A}-
\frac{\langle\heavy^{\dagger} \rangle
g^2\{T_A,T_B\}\Gamma\langle\heavy \rangle K^*_B}{M^2_A}-
\frac{\langle\heavy^{\dagger} \rangle
gT_AZ\langle\heavy \rangle}{2M^2_A}\right)\nonumber\\
&-&\frac{1}{2}\light^{\dagger} g^2\{T_A,T_B\}\light K_AK_B^*
+\light^{\dagger} gT_A\Gamma\light K_A^*\nonumber\\
&-&\frac{1}{M^2_A}\left(\langle\heavy^{\dagger} \rangle gT_A\Gamma\light K^*_B+
\light^{\dagger} gT_A\Gamma\langle\heavy \rangle K^*_B+h.c.\right)
\langle\heavy^{\dagger}\rangle g^2\{T_A,T_B\}\light+h.c.\nonumber\\
&+&\frac{1}{M^2_A}\left(\langle\heavy^{\dagger}\rangle g^2\{T_A,T_B\}\light
+h.c.\right)\left(\langle\heavy^{\dagger}\rangle g^2\{T_A,T_C\}\light+h.c.
\right)K_BK_C^*\Bigg\}\, ,
\label{fulleff}
\eea
where $K_A=\langle\heavy^{\dagger} \rangle
gT_A\Gamma\langle\heavy \rangle /M^2_A$ and
$(T_AT_BT_C)_{sym}$ is the symmetrization of the product
$T_AT_BT_C$ under the indices $A$, $B$ and $C$.

Eqs.~(\ref{triterm}) and ({\ref{fulleff})
together with eqs.~(\ref{muterm}) [from
integrating out the heavy chiral sector] and
(\ref{induced})  [from integrating out
the heavy  MSSM-non-singlet vectors] give the full
low-energy effective theory  of SSB terms
below $M_G$.

\subsection{Case with $W'(\heavy,\light)\not= aW(\heavy,\light)$}

To derive the effective Lagrangian in the previous sections,
we have assumed that
$W'\propto W$ ($W$-proportionality).
Here we want to study the implications
of relaxing such a proportionality.
This is the case of the most general softly broken supersymmetric theory.
It could also be the case that
 $W'= aW$ holds at tree-level but
 gravitational (or string \cite{louis})
corrections alter this relation (by the non-renormalization
theorem this relation cannot be modified by radiative corrections from the
GUT).

When $W'\not\propto W$,
  the mass matrix in $W'$ is not diagonal
in  the  SUSY-physical basis
 and then
terms of the order
 $M_G\heavy \light$ and  $M_G^2\heavy $
can appear in $W'$.
These terms induce new contributions to the SSB parameters of the
light superfields that have
not been considered before in the literature.
In order to reduce the number of new diagrams that can now be generated,
we can
 perform a superfield
 redefinition, ${\bf \Phi}\rightarrow {\bf \Phi}-\eta\Gamma{\bf \Phi}$,
in eq.~(\ref{softsf}) such that
the terms proportional to $\Gamma$
 can  be absorbed in $W'$
without changing $\call_{SUSY}$ \cite{yamada}\footnote{We  have not
done such a redefinition in the previous sections, since we
wanted to maintain the
proportionality between $W$ and  $W'$. Of course, this redefinition had
changed the Feynman diagrams but not the final result.}.
Also  the terms proportional to $\Lambda$  can be absorbed in $W'$.

 In fig.~4 we show the new Feynman diagrams
due to the new couplings in $W'$.
We denote by  ``$\bullet$'' a vertex
arising from $\int d^2\theta\eta W'$.
The internal line with  a ``{\bf $\times$}''  denotes a
 $\langle\heavy\heavy^{\dagger}\rangle$ propagator.
Although it goes like $\sim M_G^{-2}$, it can be compensated by
powers of $M_G$ arising from the new vertices in $\int d^2\theta\eta W'$
and give a non-negligible contribution.
The explicit contributions can be obtained from the above rules
 ({\it i})--({\it vi}) together with
\begin{quote}
\noindent ({\it vii})   For each  $\langle\heavy_i\heavy^{\dagger}_i\rangle$
 propagator,  write  $-\bar D^2/(4M^2_{ij})$.
\end{quote}
The diagrams of fig.~4a
 induce new trilinear SSB terms,
the diagrams of fig.~4b--c  induce
bilinear SSB terms and the diagrams of fig.~4d--e induce
 scalar soft masses.
In the next section
we will present some examples of models where
such diagrams are generated after integrating out the heavy modes.

\section{Phenomenological implications}

\subsection{Hierarchy destabilization}

In the absence of light MSSM-singlets
and assuming $W$-proportionality,
 the effective theory of SSB terms
is given by
eqs.~(\ref{muterm}) and
 (\ref{induced})--({\ref{fulleff}). Inspection of these terms reveals
 that there are not
 terms of ${\cal O}(M_G\msusy)$.
One can also see  this from dimensional analysis of
the diagrams of figs.~1--3.
Thus, the stability of the hierarchy, after integrating
the heavy modes, is guaranteed  for a general SSB terms if
$W'\propto W$ even in the absence of
universality.
Same conclusions have been reached in ref.~\cite{sen,murayama}
in component fields.

Nevertheless,
if $W$-proportionality does not hold at $M_P$,
we have that  new SSB terms
 are generated (such as the diagram of fig.~4b)
that can be of  $\calo(M_G\msusy)$ and  spoil the hierarchy.
One
 example where this occurs,
 is the minimal SU(5) model \cite{su5a}.
The Higgs sector of the model consists of three supermultiplets,
a Higgs fiveplet ${\bf H}$
 and antifiveplet $\overline{\bf H}$ and the adjoint
 ${\bf 24}$.
The superpotential is given by
\be
W=M_H \overline{\bf H}{\bf H}+
\lambda\, \overline{\bf H}\,{\bf24}\,{\bf  H}+W({\bf 24})\, .
\label{super}
\ee
In the supersymmetric limit
the ${\bf 24}$ develops a VEV of $\calo(M_G)$,
\be
\langle {\bf 24}\rangle =V_{24}{\bf Y}\equiv V_{24}\, {\rm diag}
(2,2,2,-3,-3)\, ,
\label{vevsigma}
\ee
that breaks SU(5) down to the MSSM gauge group. To keep light the
Higgs  SU(2)$_L$-doublets embedded in the $\overline{\bf H}$ and
 ${\bf H}$,
 we need the fine-tuning\footnote{Higher-dimension operators
suppressed by powers of $M_P^{-1}$,
 could also
be present
in W. In this case, we would  also need to fine-tune  these operators
to preserve the light Higgs of the MSSM.}
\be
M_H-3\lambda V_{24} \lesssim {\cal O}(M_Z)\, .
\label{finetuning}
\ee
However, if  the last SSB term
 of (\ref{softsf}),
\be
\int d^4\theta\eta W'({\bf \Phi})=\int d^4\theta\eta
\Big[M'_H \overline{\bf H}{\bf H}+
\lambda'\, \overline{\bf H}\,{\bf24}\,{\bf  H}+W'({\bf 24})\Big]\, ,
\label{dest}
\ee
 is present in the model
with $\lambda'\not= \lambda$ or $M_H'\not=M_H$,
a mass term of $\calo(M_G\msusy)$ for the MSSM Higgs doublets
is induced. This term is
not cancelled by the fine-tuning (\ref{finetuning}) and
 destabilize the
gauge hierarchy.

It is important to notice that the destabilization
 of the hierarchy by the
SSB parameters arises because  a heavy superfield couples
linearly to  light superfields (see diagram of fig.~4b),
 and not because
 we fine-tuned parameters.
In models without these couplings the hierarchy will be stable
even if   $W$-proportionality does not hold.
This is the case for models where the doublet-triplet
splitting is  obtained by the
missing partner or missing VEV mechanisms \cite{savas,babu}.

\subsection{$A$-terms, soft masses and FCNC}

As we discussed in section I,
deviations  from $W$-proportionality and/or eq.~(\ref{universal})
 for the squark and slepton
SSB parameters
 can have
 implications in FCNC. In the low-energy effective theory the
 SSB parameters
can be modified by ({\it 1})
renormalization effects or ({\it 2}) integrating out
the heavy modes.
Renormalization effects  can cause
$\Gamma_{\alpha\beta}$  and $Z_{\alpha\beta}$ [in
 $\call_{soft} (\heavy =0, V_A=0)$] to deviate
from  universality  if
 the two light generations have  different Yukawa couplings.
Such  contributions, however, can be suppressed if one
assumes  small Yukawa
couplings for the first and second family or a flavor symmetry.
Renormalization effects  from the top Yukawa
 can also modify the SSB parameters of the light generations
and enhance the supersymmetric-contribution to FCNC processes \cite{barbieri}.

The second type of effects (those from integrating out the
heavy modes)
can arise from  the rotation
of the superfields to the  SUSY-physical basis
and/or from  the diagrams  of fig.~1--4.
The first ones are  analyzed in ref.~\cite{us,murayamab}.
Here we will analyze the effects from diagrams  of fig.~1--4;
 their
explicit expressions are
given in eqs.~(\ref{muterm}) and
 (\ref{induced})--({\ref{fulleff}).
For the case $W'\propto W$, only
 eq.~(\ref{triterm}) can induce a trilinear SSB term,
\be
K_A\int d^4\theta\bar\eta\light^{\dagger} T_A\light\rightarrow
 K_A\phi^*_\alpha(T_A)_{\alpha\beta}\left(\frac{\partial
 W_{eff}(\light)}{\partial\phi_\beta}\right)^*\, ,
\label{trili}
\ee
which, as mentioned before,
 arises only if $T_A$ commutes with the unbroken GUT-generators.
However, from the gauge invariance of $W$ one has
\be
\light^*_\alpha(T_A)_{\alpha\beta}\left(\frac{\partial
 W}{\partial\light_\beta}\right)^*+
\light^*_\alpha(T_A)_{\alpha i}\left(\frac{\partial
 W}{\partial\heavy_i}\right)^*=0\, ,
\ee
and then  if $(T_A)_{\alpha i}=0$, \eg,
 $T_A$ is a diagonal generator, the  contribution from
 eq.~(\ref{trili}) vanish.
Therefore,
no trilinear SSB parameter is induced from
 the diagrams of fig.~1-3
in GUTs such as SU(5) and  SO(10)
which do not contain   broken generators
 that are singlet under the MSSM
group and non-diagonal.
This leads to a surprising consequence.
Consider a GUT, such as  the Georgi-Jarlskog
model \cite{georgi}, in which
 the different families couple above $M_G$ to Higgs with different
gauge quantum numbers.
As we said in  section I, one would expect that in
such theories
 the {\it gauge}  corrections  spoil the proportionality
$A_{ijk}\propto Y_{ijk}$.
Nevertheless, these effects decouple from the low-energy effective theory;
the $A$-terms for the light fields
arise only from
 $\call_{soft} (\heavy =0, V_A=0)$.
For example, in  the MSSM the trilinear term for
 $Q_\alpha U_\beta H$ is given by
\be
A_{ Q_\alpha U_\beta H}=\msusy\left[aY_{Q_\alpha U_\beta H}+
Y_{Q_\gamma U_\beta H}\Gamma_{Q_\gamma Q_\alpha}
+Y_{Q_\alpha U_\gamma H}\Gamma_{U_\gamma U_\beta}+
Y_{Q_\alpha U_\beta H}\Gamma_{HH}\right]\, ,
\ee
independently of the physics above $M_G$.
In the MSSM only  one Higgs couples to the $Q$ and $U$,
 and then   corrections to $\Gamma_{HH}$
are universal in flavor space. Note that the breaking
of universality in $\Gamma_{Q_\gamma Q_\alpha}$ and
$\Gamma_{U_\gamma U_\beta}$ can only arise from Yukawa
corrections but not from gauge corrections.

For the
light scalar  soft-masses, we have contributions from
eqs.~(\ref{triterm}) and (\ref{fulleff}).
In particular, the  terms of (\ref{fulleff})
proportional to $\light^{\dagger} T_A\light$
 (diagrams of fig.~3c)
are the so-called $D$-term contributions analyzed in ref.~\cite{dterms}.
These terms can be dangerous since they can split the
squark masses if the
 three generations
have different transformation properties under the broken generators.
This is the case in most of
the flavor theories based on a horizontal gauge symmetry.

Let us turn to the case
 $W'\not\propto W$. From    eq.~(\ref{SSBcomp}), we see that
now  $A_{ijk}$  are  not
 proportional to  $Y_{ijk}$. Thus,
large
 deviations from $W$-proportionality are
not allowed by FCNC constraints.
It is important to notice that
 even if
the breaking of $W$-proportionality
arises only in the bilinear terms ($B_{ij}\not\propto M_{ij}$)
of the heavy fields,
 diagrams like those of
 fig.~4a can induce a breaking of proportionality in the trilinear terms
 and lead to FCNC.
An example  of  a model where this can occur
 is the Georgi-Jarlskog
model \cite{georgi}. In this model the
 quarks and leptons couple to different
Higgs representations such that below $M_G$,
  only two linear combinations of them
are light (the MSSM Higgs doublets).  Since the rotation that
diagonalize the Higgs mass matrix in $W$ does not diagonalize
the mass matrix in $W'$,  mixing mass terms  of $\calo(M_G)$
between the heavy and
the light Higgs  will be present in $W'$. These terms
will induce  diagrams like those in    fig.~4a, breaking
 the proportionality $A_{ijk}\propto Y_{ijk}$
and inducing  dangerous flavor violations.

\subsection{Unification of soft masses  at $M_G$}

It has been recently shown in ref.~\cite{us}
 that GUT-relations such as eq.~(\ref{GUTrelation})  in SU(5)
can be altered  if the light quarks and leptons
 come from  different
linear combinations of a pair of  GUT-multiplets.
In our formalism this effect can be understood as follows.
If
the rotation (\ref{separation}) depends on the VEVs of the
fields that break the GUT-group, the matrices $Z$ and $\Gamma$
 after such a rotation
do not have to preserve the GUT-symmetry. Thus,
fields  embedded in  a same GUT-representation can
have different
 soft masses.  Of course, if $\Gamma,Z\propto {\bf 1}$, the rotation
(\ref{separation}) will not change $\Gamma, Z$ and  universality will be
maintained.

In the SUSY-physical basis,  there are also  effects
arising from the tree-level diagrams of fig.~2--4
that can induce  mass-splittings.
If $W'\propto W$, the only soft mass term induced in SU(5) (or in any
GUTs that does not contain
 heavy MSSM-singlet vectors) is
eq.~(\ref{induced}) -- diagram of  fig.~2.
Note, however,  that this contribution is non-zero only if
the coupling $\eta\langle\heavy^{\dagger} \rangle T_A\Gamma\light$ exists.
This coupling is present if
 the GUT-multiplet that contains  light superfields
mixes at $M_G$ with  a  GUT-multiplet
that
gets a VEV of $\calo(M_G)$.
 This condition is not satisfied in
simple models. In the minimal SU(5) model, the matter fields are in
the $\bar{\bf 5}$ and
{\bf 10} representations and cannot mix
with the {\bf 24}  that contains the MSSM-singlet.
Thus,
the absence of the term (\ref{induced}) is guaranteed and the
relation (\ref{GUTrelation}) is not
modified.
 A different situation can occur for
a SU(5)$\times$U(1) model  where
\begin{equation}
\{Q,D,\nu\}\, \in\ {\bf 10}\ ,\
\{L,U\}\, \in\ {\bf 5}\ ,\
E\,\in\ {\bf 1}\, .
\end{equation}
Now, the {\bf 10} contains a singlet under the MSSM, the
neutrino $\nu$,
 and   for $\langle\nu\rangle\sim M_G$ the diagram of fig.~2
 can induce  different  soft masses for the $Q$ and $D$.

In SO(10) models, since there is a broken U(1)-generator that
commutes with the MSSM group generators,
 the diagrams of    fig.~3
can induce shifts in the soft masses of the fields embedded
in a single SO(10) representation.
Note that only part of these contributions, diagrams of
 fig.~3c,
is usually considered
in the literature ($D$-term contribution\cite{dterms}).

If $W'$ is not proportional to $W$, extra
contributions to the soft masses come from the
diagrams of fig.~4d.
In the minimal SU(5), such diagrams do not arise since there are not
 couplings of
a light matter superfield to two heavy superfields.
In non-minimal SU(5), however, these couplings can be present
and induce the diagrams of  fig.~4d.
We will study these contributions
in the context of the model in ref.~\cite{us}.
The model is a
 SU(5) GUT  where the matter
content is extended with an extra  ${\bf 5}$ and $\bar{\bf 5}$.
The superpotential (for only one light generation) is given by
\be
W={\bf 5}_H\left[M\,\bar{\bf 5}_1+\lambda\, {\bf 24}\,\bar{\bf 5}_2
\right]+h\,{\bf 10} \overline{\bf H}  \bar{\bf 5}_1\, .
\label{superpotential}
\ee
In  the supersymmetric limit,  the ${\bf 24}$ gets a VEV given by
eq.~(\ref{vevsigma}).
One linear combination of
 $\bar{\bf 5}_1$ and
 $\bar{\bf 5}_2$ will acquire a large mass of order $\sim
M_{G}$ and the orthogonal combination will be the light quarks and leptons.
 Because the
hypercharges of  the quark and lepton embedded in the $\bar{\bf 5}$s
are different,  they will be  {\it different}
linear combinations of the corresponding states in  $\bar{\bf 5}_1$  and
$\bar{\bf 5}_2$:
\be\left(\matrix{D\cr L}\right)=-\sin\theta_Y \bar{\bf 5}_1
+\cos\theta_Y \bar{\bf 5}_2\, ,
\label{rotation}
\ee
where
\be
\sin\theta_Y=\frac{\rho{\bf Y}}{\sqrt{1+\rho^2 {\bf Y}^2}}\, ,
\ee
with $\rho=\lambda V_{24}/M$.
In this model
the diagrams of fig.~4d
 lead to extra contributions to
the soft masses of the $L$ and $D$.
For example, the first diagram of  fig.~4d induce a
soft mass
\bea
m^2_{ D}&\sim&\msusy^2 \frac{|-M's_{D}+\lambda'V_{24}c_{D}|^2}
{M^2+\lambda^2V_{24}^2}\, ,\nonumber\\
m^2_{L}&\sim&
 \msusy^2 \frac{|-M's_{L}+\lambda'V_{24}c_{L}|^2}
{M^2+\lambda^2V_{24}^2}\, ,
\eea
where $s_a$  is given by
\be
s_a=\sin\theta_{Y_a}=\frac{\rho{ Y_a}}{\sqrt{1+\rho^2 {Y_a^2}}}\, ,
\ee
and $Y_a$ is the hypercharge of $a$.
Since squarks and sleptons have different hypercharge, these contributions
break the degeneracy of their masses. These are  extra contributions to
those calculated in ref.~\cite{us}. They   go to zero
in the limit
$M'\rightarrow M$ and $\lambda'\rightarrow\lambda$ ($W$-proportionality).

\section{Conclusion}

The SSB parameters, if  generated above $M_G$, can
provide us direct information about the physics at high-energy scales.
In this paper, we have calculated the shifts induced in
 the SSB parameters of the effective theory  when the heavy modes,
coming from a GUT or a flavor theory at a high scale $M_G$,
 are integrated out at tree-level\footnote{One-loop corrections
from the heavy modes
could also
 be important for large couplings \cite{nir}.}.
 We have considered the most general softly
broken  theory and   worked within the superfield formalism. This formalism
is very suitable
 for this purpose and   the calculations  can
be easily done using
  Feynman diagrams.
For models where supersymmetry is broken by a hidden sector,
and  therefore $W$-proportionality
[eq.~(\ref{proport})] holds, the contributions to the SSB parameters of
the MSSM are given by the diagrams of fig.~1--3 and can be easily
calculated following
 the Feynman rules ({\it i})-({\it vi}) given in section II.
Some
general statements can be inferred from the above analysis.
For example, in SU(5) models we have found that the induced bilinear SSB terms
are always of $\calo(\msusy^2)$ [never of  $\calo(M_G\msusy)$] and
 neither $A$-terms
nor soft masses are generated.
Nevertheless,
mass-splittings can arise when the superfields are
rotated to the physical basis \cite{us}.

The pattern of the SSB parameters
depends strongly on  the $W$-proportionality condition.
 When it
is relaxed,  extra contributions to the $A$-terms
and soft masses are induced.
 These contributions
depend on  ratios between   VEVs  and  masses of the heavy superfields.
Hence a hierarchy of soft masses can be generated
in the low-energy spectrum even if
there is an unique  supersymmetric scale $\msusy$.
Looking at  complete GUT models
 (where the usual GUT problems,
 such as the doublet-triplet
splitting or the fermion mass spectrum, are addressed)  \cite{babu},
one finds that
most of these contributions are present and
can cause serious problems with flavor violations.
The pattern of SSB parameters at $M_G$, after integrating out the heavy states,
can be completely different from that induced at $M_P$.

\acknowledgments
We thank E. Roulet for discussions about the results of ref.~\cite{giudice}.

\pagebreak

\centerline{\large FIGURE CAPTIONS}

FIG.~1: {Tree-level diagrams generating  a
 $D$-term for the light superfields
from interchange of a heavy chiral-superfield.}

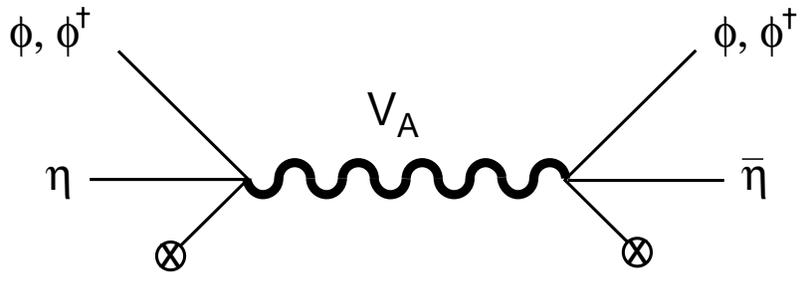
FIG.~2: {Tree-level diagram generating  a $D$-term for the light superfields
from interchange of a heavy gauge vector-superfield non-singlet
 under the MSSM.}

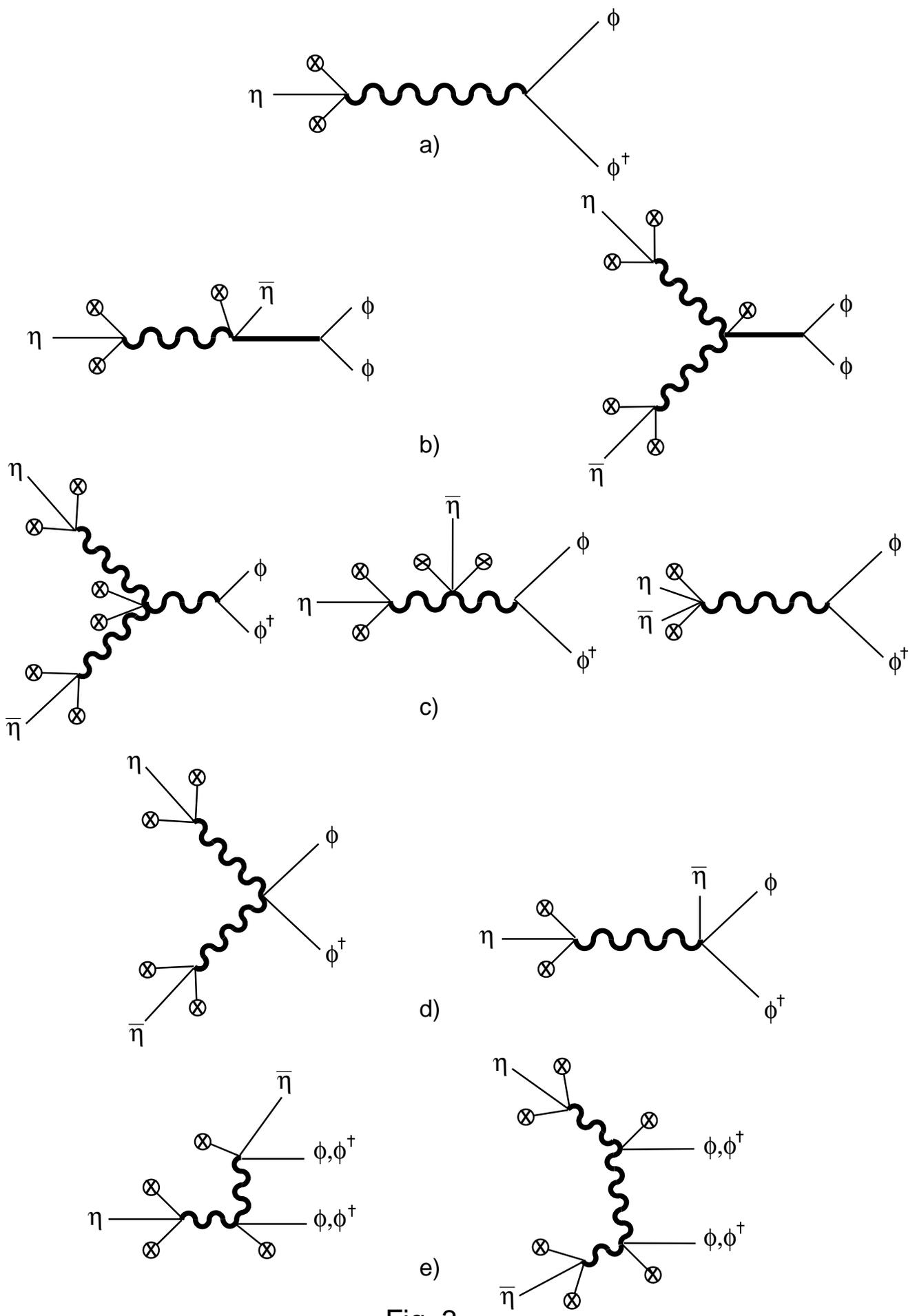
FIG.~3: {Tree-level diagrams generating a $D$-term for the light superfields
from interchange of  heavy gauge vector-superfields.}

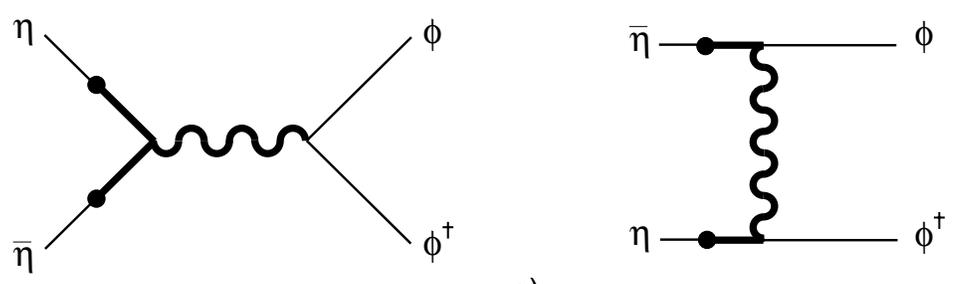
FIG.~4: {Tree-level diagrams contributing to a  $F$-term (fig.~a--d)
and D-term (fig.~e) for the light
superfields for the case $W'\not= aW$. We denote by an internal
line with   ``{\bf $\times$}''
  a $\langle\heavy\heavy^\dagger\rangle$ propagator
 and denote by  ``$\bullet$'' a vertex
 arising from $\int d^2\theta\eta W'$.}

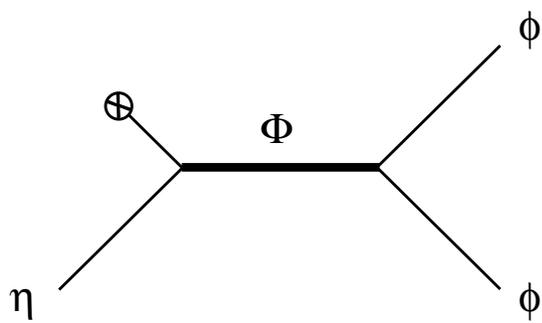
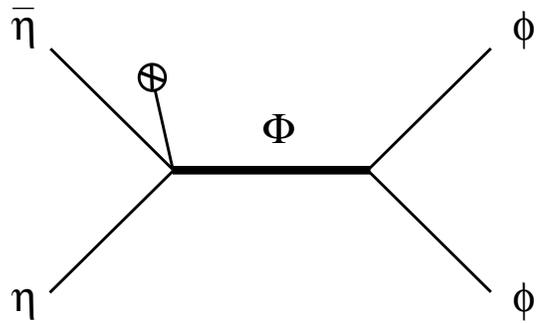
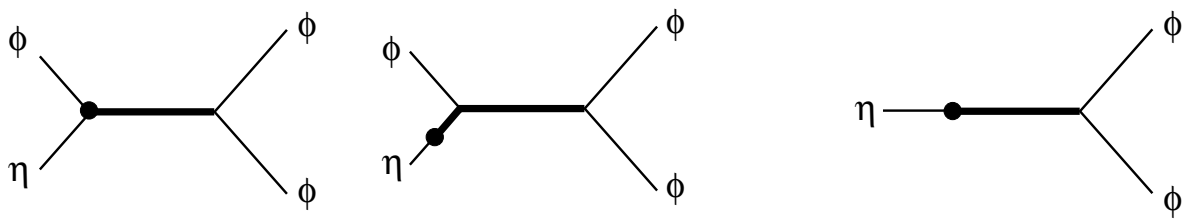
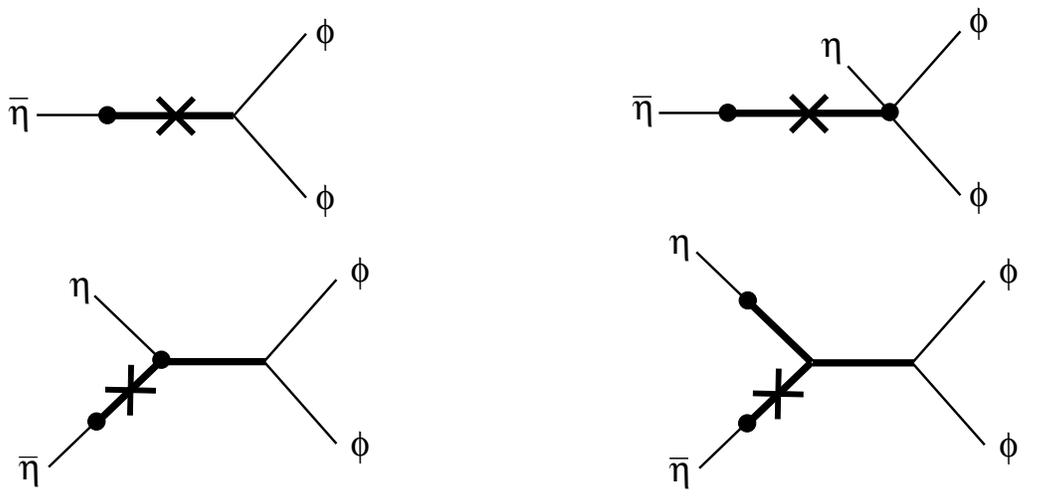
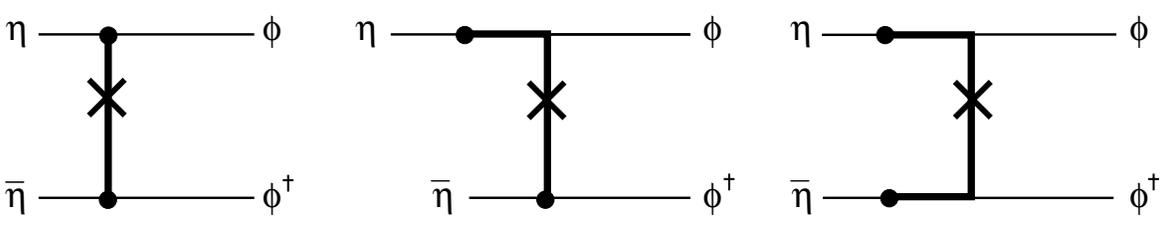
\end{document}